\documentclass[conference]{IEEEtran}
\IEEEoverridecommandlockouts
\usepackage{cite}
\usepackage{amsmath,amssymb,amsfonts}
\usepackage{algorithmic}
\usepackage{todonotes}
\usepackage{graphicx}
\usepackage{textcomp}
\usepackage{xcolor}
\def\BibTeX{{\rm B\kern-.05em{\sc i\kern-.025em b}\kern-.08em
    T\kern-.1667em\lower.7ex\hbox{E}\kern-.125emX}}

\usepackage{hyperref}
\usepackage{setspace}

\makeatletter
\def\@IEEEpubidpullup{8\baselineskip}
\makeatother
\begin{document}

\IEEEoverridecommandlockouts
\IEEEpubid{
\parbox{\columnwidth}{\vspace{-4\baselineskip}Permission to make digital or hard copies of all or
part of this work for personal or classroom use is granted without fee provided that copies are not
made or distributed for profit or commercial advantage and that copies bear this notice and the full
citation on the first page. Copyrights for components of this work owned by others than ACM must
be honored. Abstracting with credit is permitted. To copy otherwise, or republish, to post on servers
or to redistribute to lists, requires prior specific permission and/or a fee. Request permissions from
\href{mailto:permissions@acm.org}{permissions@acm.org}.\hfill\vspace{-0.8\baselineskip}\\
\begin{spacing}{1.2}
\small\textit{ASONAM '19}, August 27-30, 2019, Vancouver, Canada \\
\copyright\space2019 Association for Computing Machinery. \\
ACM ISBN 978-1-4503-6868-1/19/08 \\
\url{http://dx.doi.org/10.1145/3341161.3342954}
\end{spacing}
\hfill}
\hspace{0.9\columnsep}\makebox[\columnwidth]{\hfill}}
\IEEEpubidadjcol

\title{Exploring the Context of Course Rankings on Online Academic Forums}

\author{\IEEEauthorblockN{Taha Hassan, Bob Edmison}
\IEEEauthorblockA{Computer Science Department \\
Virginia Tech\\
Blacksburg, VA \\
\{taha, kedmison\}@vt.edu}
\and
\IEEEauthorblockN{Larry Cox II, Matthew Louvet, Daron Williams}
\IEEEauthorblockA{Technology-enhanced Learning and Online Strategies \\
Virginia Tech\\
Blacksburg, VA \\
\{lacox, mattl06, debo9\}@vt.edu}
}

\maketitle

\begin{abstract}
University students routinely use the tools provided by online course ranking forums to share and discuss their satisfaction with the quality of instruction and content in a wide variety of courses. Student perception of the efficacy of pedagogies employed in a course is a reflection of a multitude of decisions by professors, instructional designers and university administrators. This complexity has motivated a large body of research on the utility, reliability and behavioral correlates of course rankings. There is, however, little investigation of the (potential) implicit student bias on these forums towards desirable course outcomes at the institution level. To that end, we examine the connection between course outcomes (student-reported GPA) and the overall ranking of the primary course instructor, as well as rating disparity by nature of course outcomes, based on data from two popular academic rating forums. Our experiments with ranking data about over ten thousand courses taught at Virginia Tech and its 25 SCHEV-approved peer institutions indicate that there is a discernible albeit complex bias towards course outcomes in the professor ratings registered by students.
\end{abstract}


\section{Introduction}
Online forums for ranking course instructors, like \textit{RateMyProfessors} \cite{rmphomepage} and \textit{Koofers} \cite{koofershomepage} are a popular resource among university students for detailing their perception of the quality of course instruction and content. These forums cater to a number of their contributors' needs, including but not limited to information seeking, gratification, and convenience \cite{kindred2005he}. They have therefore, inspired considerable research attention over the years \cite{chang2014koofers} \cite{brown2009rating} \cite{legg2012ratemyprofessors} \cite{kindred2005he}. Prior studies have identified several broad themes in the student feedback sampled from these forums, including teacher personality, aptitude and preparation, ease of access to help and feedback from course staff, and perceived practicality of the course rubric \cite{hartman2013ratemyprofessors}. There is however, lesser attention devoted in this literature, to institution-level correlates of student perception. Empirical investigations of the potential sources of student bias in said perception are often divided in their conclusions, because of limitations of sample sizes or meta-variable space \cite{marsh1984} \cite{centragradestudy} \cite{feldman2007review}. Assessing its reliability at scale can lend insights to instructional designers, department administrators and instructors alike on the limitations of existing pedagogies. It can also potentially extend the utility of university-managed end-of-semester course evaluations, and help improve the usability, relevance, accessibility and trustworthiness \cite{legg2012ratemyprofessors}\cite{hassan2019trust} of its host forums.

\begin{figure}
\includegraphics[width=0.5\textwidth]{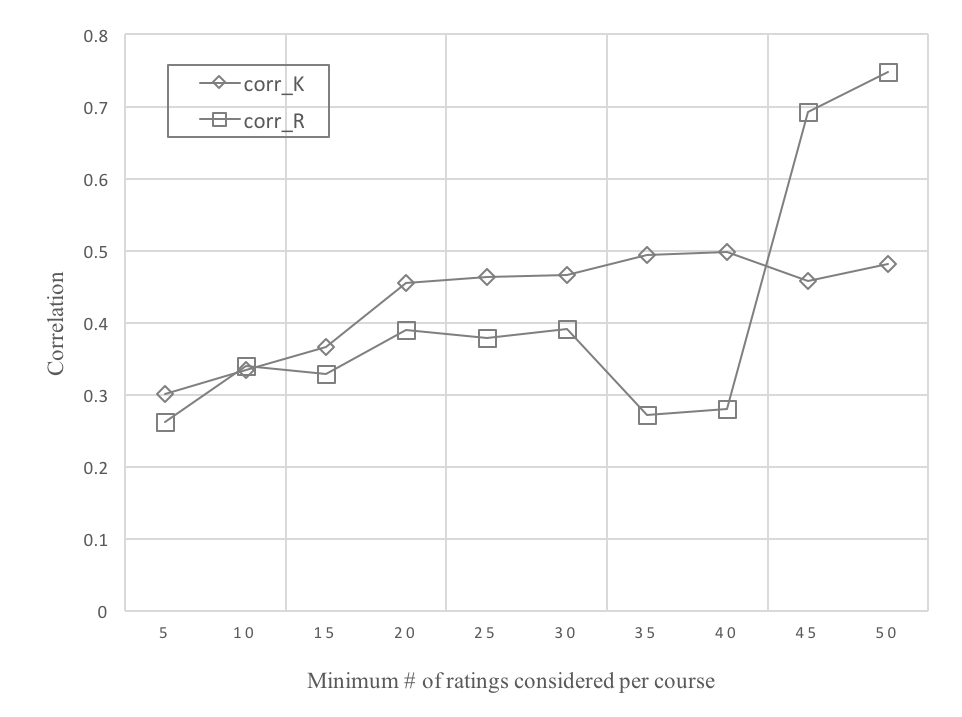}
\caption{Correlation between the average student-reported GPA and overall instructor rating, function of the minimum number of ratings considered towards a course (\textbf{R}: RateMyProfessors \cite{rmphomepage}, \textbf{K}: Koofers \cite{koofershomepage})}
\label{fig:gap}
\end{figure} 

\begin{table*}[t]
\caption{Key counts for courses in the historical ratings dataset from RateMyProfessors (\textbf{R}) and Koofers (\textbf{K})}
\centering
\label{tab_key_counts}
\begin{tabular}{|p{2.2cm}| p{1.6cm}| p{2.2cm}| p{1.6cm}| p{1.6cm}| p{2.2cm}| p{1.6cm}|}
\hline
\bfseries Institution & \bfseries Courses (R) & \bfseries Departments (R) & \bfseries Ratings (R) & \bfseries Courses (K) & \bfseries Departments (K) & \bfseries Ratings (K)\\
\hline
ISU & 742 & 141 & 63192 & - & - & -\\
\hline
UC-D & 1055 & 115 & 57986 & - & - & -\\
\hline
PSU & 962 & 148 & 50786 & - & - & -\\
\hline
UFL & 615 & 176 & 39296 & 149 & 85 & 4359\\
\hline
UM & 704 & 139 & 33606 & - & - & -\\
\hline
UWISC-M & 646 & 159 & 32467 & - & - & -\\
\hline
UMN-TC & 718 & 128 & 31265 & - & - & -\\
\hline
SB & 519 & 84 & 26139 & - & - & -\\
\hline
PITT & 603 & 112 & 25566 & - & - & -\\
\hline
SUNY-B & 418 & 88 & 24849 & - & - & -\\
\hline
TAMU & 365 & 99 & 22043 & 154 & 56 & 6003\\
\hline
UIUC & 482 & 99 & 21350 & 126 & 51 & 3569\\
\hline
UCB & 488 & 117 & 20519 & - & - & -\\
\hline
USC & 473 & 86 & 18692 & - & - & -\\
\hline
UTA & 328 & 79 & 17154 & 57 & 25 & 1910\\
\hline
PURDUE & 433 & 78 & 15831 & - & - & -\\
\hline
UMD & 276 & 64 & 11806 & 254 & 64 & 11581\\
\hline
UCB & 277 & 64 & 9454 & - & - & -\\
\hline
UMC & 216 & 76 & 8328 & 75 & 34 & 2000\\
\hline
NCSU & 162 & 42 & 7873 & 61 & 29 & 1923\\
\hline
RUTGERS & 204 & 63 & 6790 & - & - & -\\
\hline
VT & 99 & 36 & 3254 & 597 & 80 & 42503\\
\hline
UW-S & 76 & 22 & 1718 & - & - & -\\
\hline
MSU & - & - & - & 888 & 110 & 67086\\

\hline
\end{tabular}
\end{table*}

\section{Related Work}
A number of studies in the educational psychology literature have examined the correlation between course outcomes and student evaluations, albeit for individual students or a course \cite{stumpf_apa}  \cite{feldman1989association} \cite{feldman2007review}. These studies generally report this correlation to be modest, somewhere between +0.1 and +0.3 \cite{feldman2007review}. Our preliminary inquiry at the institution level demonstrates that this correlation often matches and sometimes exceeds said figures \cite{hassan2019bias}. The aforementioned work notes that given the learning acquired by students during the course of a given class or academic term, all of this observed correlation can not necessarily be a result of implicit, time-invariant student bias towards course outcomes. The multi-faceted nature of student perception can affect this connection. This complexity is echoed in a comparative study of in-class assessments, and pre- and post-assessment ratings on RateMyProfessors \cite{legg2012ratemyprofessors}. The study reported how pre-assessment course ratings on instructor clarity were significantly lower than both in-class and post-assessment reviews. However, instructor easiness was reviewed lower in-class relative to online. Our study attempts to initiate a line of large-scale contextual inquiry of these ratings across institutions that can potentially help consolidate these differing interpretations.

\section{Approach}
We pursue preliminary evidence of what appears to be a modest to strong relative connection between aggregate course outcomes and student perception of the course instructor. Figure \ref{fig:gap} visualizes the correlation between these two for courses taught at Virginia Tech (597 and 99 courses rated on \textit{Koofers} and \textit{RateMyProfessors}, respectively). For the Koofers dataset, as we increase the minimum number of ratings per course considered towards the correlation, this correlation increases and achieves a steady average value of about 0.47 beyond 25 ratings per course. We explore this further by examining the disparity of instructor ranking between high, medium and low GPA student groups, as well as regressing student approval against course outcomes and perceived difficulty of various course instruments.

\begin{table*}[ht]
\caption{Hypothesis-testing the relationship b/w course outcomes and instructor rankings}
\centering
\label{tab_hyp}
\begin{tabular}{|p{2.8cm}| p{2.5cm}| p{2.5cm}| p{1.4cm}| p{1.4cm}| p{1.4cm}| p{1.4cm}|}
\hline
\bfseries Institution & \bfseries $Corr, p$ & $F, p$ & $\mu_{ov}$ & $\mu_{hi}$ & $\mu_{med}$ & $\mu_{low}$\\
\hline
ISU & 0.298, 9.5e-17* & 28.1, 1.6e-12$^{\dagger}$ & 3.71 & 3.91 & 3.66 & 3.33\\
\hline
UC-D & 0.206, 1.3e-11* & 21.8, 5.1e-10$^{\dagger}$ & 3.69 & 3.83 & 3.61 & 3.5\\
\hline
PSU & 0.239, 5.2e-14* & 32.2, 2.8e-14$^{\dagger}$ & 3.62 & 3.77 & 3.51 & 3.23\\
\hline
UFL & 0.355, 9.1e-20* & 44.4, 9.3e-19$^{\dagger}$ & 3.75 & 3.95 & 3.52 & 3.12\\
\hline
UFL (\textbf{K}) & 0.32, 6.8e-5* & 3.11, 0.04$^{\dagger}$ & 4.06 & 4.31 & 4.04 & 3.73\\
\hline
UM & 0.15, 5.5e-5* & 12.7, 3.7e-6$^{\dagger}$ & 3.8 & 3.89 & 3.65 & 3.76\\
\hline
UWISC-M & 0.245, 2.4e-10* & 24.3, 6.4e-11$^{\dagger}$ & 3.63 & 3.8 & 3.45 & 3.52\\
\hline
UMN-TC & 0.295, 6.7e-16* & 28.1, 1.72e-12$^{\dagger}$ & 3.62 & 3.8 & 3.46 & 3.19\\
\hline
SB & 0.254, 4e-9* & 17, 6.8e-8$^{\dagger}$ & 3.65 & 3.82 & 3.51 & 3.35\\
\hline
PITT & 0.294, 1.6e-13* & 31.1, 1.3e-13$^{\dagger}$ & 3.72 & 3.9 & 3.54 & 3.51\\
\hline
SUNY-B & 0.362, 2e-14* & 31.3, 2e-13$^{\dagger}$ & 3.61 & 3.88 & 3.43& 2.98\\
\hline
TAMU & 0.605, 6.9e-38* & 70.6, 1.2e-26$^{\dagger}$ & 3.74 & 4.15 & 3.53 & 2.5\\
\hline
TAMU (\textbf{K}) & 0.492, 8.5e-11* & 13.4, 4.3e-6$^{\dagger}$ & 3.89 & 4.44 & 3.94 & 3.64\\
\hline
UIUC & 0.269, 1.7e-9* & 18.1, 2.4e-8$^{\dagger}$ & 3.63 & 3.76 & 3.49 & 2.94\\
\hline
UIUC (\textbf{K}) & 0.224, 0.01* & 3.51, 0.03$^{\dagger}$ & 3.63 & 3.96 & 3.66 & 3.3\\
\hline
UCB & 0.178, 7e-5* & 11.4, 1.4e-5$^{\dagger}$ & 3.76 & 3.85 & 3.64 & 3.36\\
\hline
USC & 0.366, 1.9e-16* & 25.8, 2.1e-11$^{\dagger}$ & 3.72 & 3.88 & 3.45 & 2.93\\
\hline
UTA & 0.486, 7.2e-21* & 25.8, 3.7e-11$^{\dagger}$ & 3.78 & 4.04 & 3.57 & 3.56\\
\hline
UTA (\textbf{K}) & 0.312, 0.01* & 1.09, 0.34 & 3.75 & 4 & 3.81 & 3.59\\
\hline
PURDUE & 0.268, 1.3e-8* & 9.96, 5.9e-5$^{\dagger}$ & 3.55 & 3.7 & 3.47 & 2.97\\
\hline
UMD & 0.28, 2.1e-6* & 8.7, 2e-4$^{\dagger}$ & 3.61 & 3.81 & 3.48 & 3.52\\
\hline
UMD (\textbf{K}) & 0.354, 6.6e-9* & 5.68, 3e-3$^{\dagger}$ & 3.72 & 412 & 3.72 & 3.42\\
\hline
UCB & 0.363, 4.5e-10* & 14.3, 1.1e-6$^{\dagger}$ & 3.61 & 3.9 & 3.49 & 3.3\\
\hline
UMC & 0.399, 1.1e-9* & 9.9, 7.3e-5$^{\dagger}$ & 3.56 & 3.82 & 3.44 & 3.21\\
\hline
UMC (\textbf{K}) & 0.342, 2e-3* & 4.23, 0.02$^{\dagger}$ & 3.61 & 4.92 & 3.66 & 3.47\\
\hline
NCSU & 0.352, 4.2e-6* & 12.1, 1.2e-5$^{\dagger}$ & 3.67 & 3.89 & 3.53 & 3.55\\
\hline
NCSU (\textbf{K}) & 0.361, 4e-3* & 4.3, 0.01$^{\dagger}$ & 3.71 & 4.43 & 3.75 & 3.53\\
\hline
RUTGERS & 0.184, 8e-3* & 3.3, 0.03$^{\dagger}$ & 3.49 & 3.59 & 3.49 & 3.18\\
\hline
VT & 0.34, 5e-4* & 10.3, 8.1e-5$^{\dagger}$ & 3.63 & 3.99 & 3.46 & 3.35\\
\hline
VT (\textbf{K}) & 0.33, 4e-17* & 15.3, 3.1e-7$^{\dagger}$ & 3.83 & 4.15 & 3.86 & 3.52\\
\hline
UW & 0.102, 0.37 & 0.83, 0.43 & 3.69 & 3.77 & 3.62 & 4\\
\hline
MSU (\textbf{K}) & 0.241, 3.1e-13* & 14.7, 5e-7$^{\dagger}$ & 3.82 & 3.88 & 3.85 & 3.46\\
\hline

\multicolumn{4}{l}{$^{*}$stat. significant, $\alpha=0.05$}\\
\multicolumn{4}{l}{$^{\dagger}$stat. significant, $F>F_{crit}, \alpha=0.05$}\\
\end{tabular}
\end{table*}

\begin{table}[ht]
\caption{Regression analysis: overall professor rating, function of student GPA (\textbf{X1}), and perceived ease of course instruments (\textbf{X2} - \textbf{X5})}
\centering
\label{tab_reg}
\begin{tabular}{|p{2.2cm}| p{1.2cm}| p{1.2cm}| p{0.8cm}| p{0.8cm}|}
\hline
 & \textbf{coef} & \textbf{s. error} & $t$ & $p$\\
\hline
\textbf{intercept} & 2.6 & 0.318 & 8.3 & 0.00*\\
\hline
\textbf{X1}: GPA & 0.5 & 0.085 & 6.0 & 0.00*\\
\hline
\textbf{X2}: exams & -0.03 & 0.041 & -0.8 & 0.398\\
\hline
\textbf{X3}: quizzes & -0.08 & 0.037 & -2.3 & 0.02*\\
\hline
\textbf{X4}: projects & -0.04 & 0.027 & -1.7 & 0.08\\
\hline
\textbf{X5}: homework & -6e-3 & 0.032 & -0.19 & 0.84\\
\hline
\end{tabular}
\end{table}

\begin{table}[ht]
\caption{Regression analysis (cont.)}
\centering
\label{tab_reg2}
\begin{tabular}{|p{3.0cm}| p{1.5cm}|}
\hline
\textbf{stat.} & \textbf{val.}\\
\hline
R-squared & 0.193\\
\hline
Adj R-squared & 0.182\\
\hline
F-statistic & 17.8\\
\hline
Prob (F-statistic) & 8e-16*\\
\hline
Log-likelihood & -244.4\\
\hline
AIC & 500.8\\
\hline
BIC & 524.4\\
\hline
\end{tabular}
\end{table}

\section{Evaluation}
\subsection{Datasets}
We leverage a set of peer institutions for Virginia Tech, identified by State Council of Higher Education for Virginia (SCHEV) \cite{peers}, based on criteria such as enrollments, research funding, degrees awarded and the Carnegie Classification \cite{shulman2001carnegie}. We scraped course metadata from RateMyProfessors \cite{rmphomepage} and Koofers \cite{koofershomepage}, two popular forums for sharing course content and instructor reviews. For a university, we consider all courses with a minimum of 10 ratings. A course can have a multitude of instructors and offerings. Instructor ratings on Koofers and RateMyProfessors both, are on a 0 to 5 scale while GPA reports are on a 4.0 scale. We define the minimum acceptable use of the forum (at the institution level) as an excess of 1000 total ratings, with the course count at least 1.5X - 2X that of the department count. 

\subsection{Methods}
We significance-test the disparity in professor ratings by GPA groups using one-way ANOVA (F-test, table \ref{tab_hyp}). We also use the Python package \textit{statsmodels.OLS} towards regression analysis of instructor ratings as a function of student GPA, as well as their self-reported ease of course instruments (exams, quizzes, projects and homeworks). 

\subsection{Results}
Table \ref{tab_hyp} lists the correlations between average professor rating and student GPA for all institutions considered, as well as average ratings for each GPA group. The institutions in the \textit{RateMyProfessors} dataset report an average correlation of 0.30, while those in the \textit{Koofers} dataset report an average of 0.27. The group disparities by course outcome are modest to strong across institutions, with an average F-statistic of 20.6 and 5.86 for the \textit{RateMyProfessors} and \textit{Koofers} datasets, respectively. TAMU, for instance, reports the strongest correlation (0.605) and effect sizes (F-statistic: 70.6). However, the modest average correlation hints at department-level and course-level distinctions, that partially neutralize the differences between course outcome groups, or in rare cases like UW, eliminate the effect altogether. Consider UMD, for instance, where a relatively modest effect size (F-statistic: 8.7) coincides with a larger average instructor ranking on behalf of the low-GPA group relative to the medium GPA group. In general, however, we find that a modest connection between course outcomes and student evaluations exists across institutions with remarkable consistency. Table \ref{tab_reg} and table \ref{tab_reg2} report the coefficients, errors and significance tests for regression analysis with average instructor rating as a function of average student-reported GPA \textbf{X1}, average perceived difficulty of exams, quizzes, projects and homework (\textbf{X2} through \textbf{X5}). Course outcomes outweigh the perceived ease of course evaluations in their aggregate effect on instructor ratings ($t=6.0, t=-2.3$ for \textbf{X1} and \textbf{X3}, respectively). Difficulty of quizzes appears modestly relevant in ascertaining the overall student satisfaction with the course, with higher difficulty linked to lower student approval (t-statistic is negative). 

\section{Discussion}
We intend to expand our analysis by considering the time order of the instructor ratings in our dataset. While the magnitude of the observed correlation between course outcomes and student rankings is nearly consistent across institutions (with minimum aggregate forum use) we tested for, it is harder to argue about the directionality of this correlation without said data. 
We are also working to expand our contextual variable space to include university mandate (teaching vs. research, public vs. private), course modality (STEM/non-STEM, undergraduate/graduate, in-class/online), assessments (group projects, pop quizzes, class participation), instructor attributes (accessible outside the class, competent, tendency to provide feedback), technology use (LMS and third-party apps for course management, testing and assessment), logistics (instructional design training, number of TAs, etc.), and forum features (content management and editorial control).
Another important step in realizing this contextual inquiry is designing metrics that summarize the observed disparity between effects of course rubric, content quality, interaction fidelity of host forums as well as course outcomes on the overall instructor ranking. These metrics are a crucial first step in determining the uncertainty and/or bias implicit in instructor ratings across departments and institutions.

\section{Conclusion}
Our study examines the strength of the connection between course outcomes and aggregate instructor rank on two popular academic forums. We find that for several academic institutions, student ratings of course instructors incorporate disparities linked to course outcomes as opposed to the student perception of the relative ease of course materials, content and evaluations. We intend to generalize this analysis into a robust approach of isolating and correcting for bias on academic forums.

\bibliographystyle{IEEEtran}
\bibliography{sample-bibliography}

\end{document}